\documentclass[aps,prl,twocolumn,showpacs,floatfix,superscriptaddress,citeautoscript,noshowpacs,longbibliography]{revtex4-1}
\usepackage{graphicx}
\usepackage{amsmath}
\usepackage{amssymb}
\usepackage{color}
\usepackage{float}
\usepackage[svgnames]{xcolor}
\usepackage[colorlinks=true,citecolor=blue,linkcolor=blue,pdfstartview=FitH,bookmarks=False,pdfpagemode=UseNone]{hyperref}

\begin{document}

\flushbottom

\title{Anisotropic superconductivity and Fermi surface reconstruction in the spin-vortex antiferromagnetic superconductor CaK(Fe$_{0.95}$Ni$_{0.05}$)$_4$As$_4$} 

\author{Jos\'e Benito Llorens}
\affiliation{Laboratorio de Bajas Temperaturas y Altos Campos Magn\'eticos, Departamento de F\'isica de la Materia Condensada, Instituto Nicol\'as Cabrera and Condensed Matter Physics Center (IFIMAC), Unidad Asociada UAM-CSIC, Universidad Aut\'onoma de Madrid, E-28049 Madrid,
Spain}

\author{Edwin Herrera}
\affiliation{Laboratorio de Bajas Temperaturas y Altos Campos Magn\'eticos, Departamento de F\'isica de la Materia Condensada, Instituto Nicol\'as Cabrera and Condensed Matter Physics Center (IFIMAC), Unidad Asociada UAM-CSIC, Universidad Aut\'onoma de Madrid, E-28049 Madrid,
Spain}

\author{V\'ictor Barrena}
\affiliation{Laboratorio de Bajas Temperaturas y Altos Campos Magn\'eticos, Departamento de F\'isica de la Materia Condensada, Instituto Nicol\'as Cabrera and Condensed Matter Physics Center (IFIMAC), Unidad Asociada UAM-CSIC, Universidad Aut\'onoma de Madrid, E-28049 Madrid,
Spain}

\author{Beilun Wu}
\affiliation{Laboratorio de Bajas Temperaturas y Altos Campos Magn\'eticos, Departamento de F\'isica de la Materia Condensada, Instituto Nicol\'as Cabrera and Condensed Matter Physics Center (IFIMAC), Unidad Asociada UAM-CSIC, Universidad Aut\'onoma de Madrid, E-28049 Madrid,
Spain}

\author{Niclas Heinsdorf}
\affiliation{Institut f\"ur Theoretische Physik, Goethe-Universit\"at Frankfurt, Max-von-Laue-Strasse 1, 60438 Frankfurt am Main, Germany}

\author{Vladislav Borisov}
\affiliation{Institut f\"ur Theoretische Physik, Goethe-Universit\"at Frankfurt, Max-von-Laue-Strasse 1, 60438 Frankfurt am Main, Germany}
\affiliation{Department of Physics and Astronomy,
Uppsala University, Box 516, SE-75120 Uppsala, Sweden}

\author{Roser Valent\'i}
\affiliation{Institut f\"ur Theoretische Physik, Goethe-Universit\"at Frankfurt, Max-von-Laue-Strasse 1, 60438 Frankfurt am Main, Germany}

\author{William R. Meier}
\affiliation{Ames Laboratory, Ames and Department of Physics $\&$ Astronomy, Iowa State University, Ames, IA 50011}

\author{Sergey Bud'ko}
\affiliation{Ames Laboratory, Ames and Department of Physics $\&$ Astronomy, Iowa State University, Ames, IA 50011}

\author{Paul C. Canfield}
\affiliation{Ames Laboratory, Ames and Department of Physics $\&$ Astronomy, Iowa State University, Ames, IA 50011}

\author{Isabel Guillam\'on}
\affiliation{Laboratorio de Bajas Temperaturas y Altos Campos Magn\'eticos, Departamento de F\'isica de la Materia Condensada, Instituto Nicol\'as Cabrera and Condensed Matter Physics Center (IFIMAC), Unidad Asociada UAM-CSIC, Universidad Aut\'onoma de Madrid, E-28049 Madrid,
Spain}

\author{Hermann Suderow}
\affiliation{Laboratorio de Bajas Temperaturas y Altos Campos Magn\'eticos, Departamento de F\'isica de la Materia Condensada, Instituto Nicol\'as Cabrera and Condensed Matter Physics Center (IFIMAC), Unidad Asociada UAM-CSIC, Universidad Aut\'onoma de Madrid, E-28049 Madrid,
Spain}

\begin{abstract}
High critical temperature superconductivity often occurs in systems where an
antiferromagnetic order is brought near $T=0K$ by slightly modifying pressure or doping. CaKFe$_4$As$_4$ is a superconducting, stoichiometric iron pnictide compound
showing optimal superconducting critical temperature with $T_c$ as large as
$38$ K. Doping with Ni induces a decrease in $T_c$ and the onset of spin-vortex antiferromagnetic order, which consists of spins pointing inwards to or outwards from alternating As sites on the diagonals of the in-plane square Fe lattice.
Here we study the band structure of CaK(Fe$_{0.95}$Ni$_{0.05}$)$_4$As$_4$
(T$_c$ = 10 K, T$_N$  = 50 K) using quasiparticle interference with a Scanning
Tunneling Microscope (STM) and show that the spin-vortex order induces a Fermi surface reconstruction and a fourfold superconducting gap anisotropy.  \end{abstract}

\maketitle

Iron pnictide superconductors mostly crystallize in a tetragonal structure.
Optimal $T_c$ appears in a phase diagram that shows structural, nematic or
magnetic order in the vicinity of
superconductivity\cite{doi:10.1021/ja800073m,doi:10.1146/annurev-conmatphys-070909-104041,Paglione2010,Hirschfeld_2011,HOSONO2015399}.
Whereas most Fe-based superconductors need doping (or pressure) to reach
maximal T$_c$ values, CaKFe$_4$As$_4$ is superconducting with the highest
critical temperature in the pure stoichiometric compound with $T_c\approx$ 38
K\cite{doi:10.1021/jacs.5b12571,PhysRevB.94.064501}. Elastoresistivity, nuclear magnetic resonance
(NMR) and neutron scattering experiments reveal magnetic
fluctuations\cite{PhysRevB.98.140501,PhysRevB.96.104512,doi:10.7566/JPSJ.86.093703}.
Contrary to other pnictide superconductors, there are neither structural
modifications of the crystal when cooling nor strong electronic anisotropy in
form of nematicity\cite{PhysRevB.94.064501,PhysRevMaterials.1.013401}. The
superconducting gap exchanges sign in different pockets of the Fermi surface
and has $s\pm$ symmetry as many other iron
pnictides\cite{PhysRevB.95.100502,PhysRevB.97.134501,PhysRevLett.117.277001,PhysRevB.95.100502,UMMARINO201650}.
Electron count and other physical properties such as T$_c$ and pairing symmetry
are similar to the nearly optimally doped
(Ba$_{0.5}$K$_{0.5}$)Fe$_2$As$_2$\cite{PhysRevB.95.100502,PhysRevB.98.020504},
where the magnetic order of BaFe$_2$As$_2$ is suppressed by hole doping with K.

\begin{figure*}
	\includegraphics[width=1.0\textwidth]{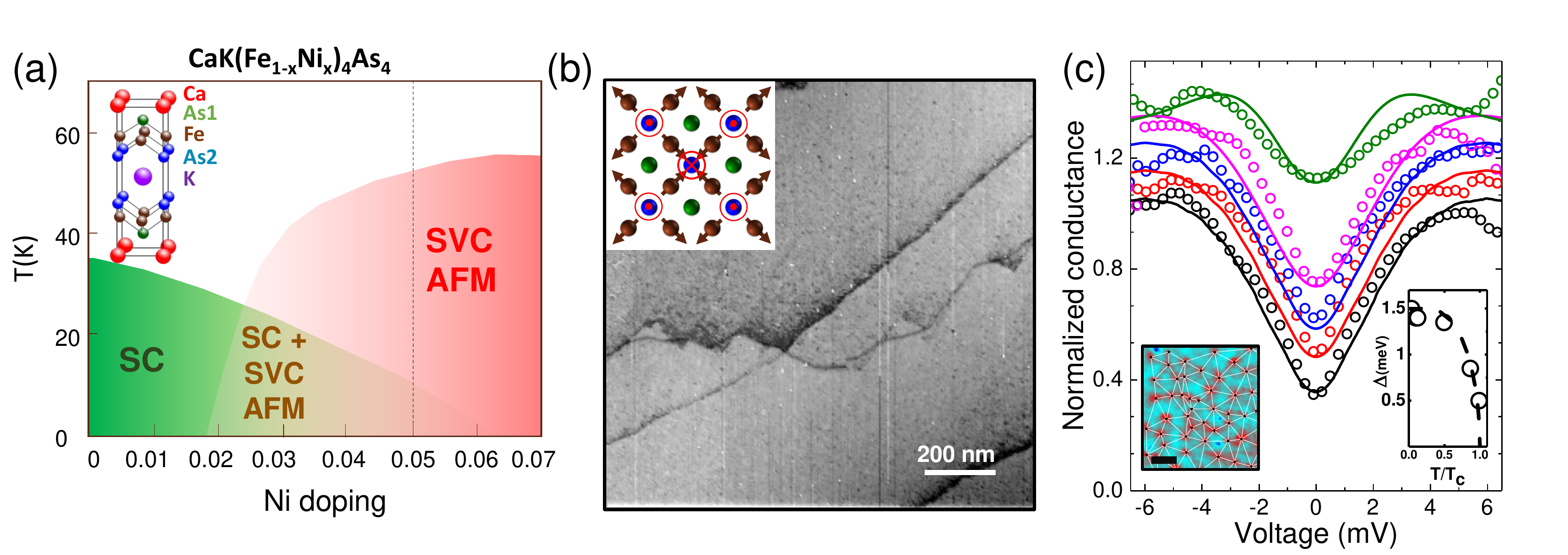}
\vskip -0.2cm
	\caption{(a) Schematic phase diagram of Ni doped CaKFe$_4$As$_4$, with a dashed vertical line indicating the Ni concentration discussed here. Crystalline structure of CaK(Fe$_{0.95}$Ni$_{0.05}$)$_4$As$_4$ is shown in the upper left inset. (b) STM topographic image of the surface of CaK(Fe$_{0.95}$Ni$_{0.05}$)$_4$As$_4$. The difference between black and white corresponds to a height change of 0.3 nm. In the inset we show a view from the top of the structure, indicating Fe (brown) and As (As1 in blue and As2 in green) atoms and with arrows indicating the spin-vortex magnetic order. Note that the magnetic moments point towards As1, giving a finite hyperfine field pointing upwards along the c-axis (red circles with a cross) and from As1 atoms, giving a hyperfine field pointing downwards along the c-axis (red circle with a dot)\cite{Meier2018}. At As2 the field cancels. (c) The temperature dependence of the tunneling conductance is shown as open circles in the main panel. Curves are taken (from bottom to top), at 0.3 K, 0.8 K, 1.4 K, 4 K and 7 K. Lines are fits using the density of states obtained for a distribution of value of the superconducting gap. The bottom right inset shows as black open circles the temperature dependence of the superconducting gap value, extracted from the maximum in the derivative of the density of states as a function of temperature normalized to $T_c$. The dashed line is a guide to the eye. Bottom left inset provides an image of the vortex lattice taken at 0.3 K and 2 T. Color scale shows the zero bias conductance which goes from the normal state value (red) to its value at zero field (blue). White lines are the Delaunay triangulation of vortex positions, which are shown as black dots. Black scale bar is 30 nm long. Further images and details are provided in Ref.\,\protect\cite{SupplementMat}.}
	\label{Fig1}
	\end{figure*}

Following this idea, electron doping by substituting Fe with Co or Ni leads to
antiferromagnetic order in CaKFe$_4$As$_4$ (Fig.\,\ref{Fig1}(a))\cite{Meier2018}. The crystal
structure is composed of Fe$_2$As$_2$ layers that are separated alternatively
with Ca and K. Thus, As sites in each layer are not equivalent, because their distance to the Fe plane differs due to being close either to Ca or to K. The distance between As1 and Ca is different than the distance between As2 and K (upper left inset in Fig.\,\ref{Fig1}(a)). This eliminates the glide symmetry in the Fe$_2$As$_2$
planes that exists in compounds such as BaFe$_2$As$_2$. As a consequence,
antiferromagnetic order is non-collinear, with spins at each of the
four Fe sites in the crystal structure pointing inwards to (or outwards from) the As sites,
giving a hedgehog spin-vortex crystal (SVC, brown arrows in upper left inset in
Fig.\,\ref{Fig1}(b))\cite{PhysRevB.93.144414,Meier2018}. The magnetic
wavevector is the same as for the usual spin density wave
antiferromagnetic or spin-charge magnetic
order\cite{PhysRevB.91.024401,PhysRevB.92.214509,PhysRevB.93.144414,ding2018hedgehog}.
This SVC order produces a characteristic pattern of hyperfine fields at the As sites depicted in the inset of Fig.\,\ref{Fig1}(b). The As1 sites have an alternating field up and down along the c axis (red circles and crosses). Critically, the hyperfine field is zero at As2 due to canceling contributions from surrounding Fe moments\cite{Meier2018,MeierPhD}. There is robust experimental evidence for the presence of the SVC within the
superconducting phase\cite{Meier2018,Ding17,bud2018coexistence,khasanov2020magnetism,kreyssig2018antiferromagnetic}.
However, the electronic band structure is yet unknown. Here we
study the local density of states of CaK(Fe$_{0.95}$Ni$_{0.05}$)$_4$As$_4$
(T$_{SVC}=50$ K and $T_c = 10$ K) via Scanning Tunneling Microscopy (STM).
We determine the band structure in the magnetic phase and show that the superconducting gap is highly anisotropic due to magnetism.

We study single crystals of CaK(Fe$_{0.95}$Ni$_{0.05}$)$_4$As$_4$ which have been obtained using the method of Ref.\,\cite{PhysRevMaterials.1.013401,Meier2018}. Samples were mounted into a dilution refrigerator STM as described in Ref.\cite{doi:10.1063/1.3567008}. We provide further details of crystals, low temperature cleaving mechanism and data analysis in Ref.\cite{SupplementMat}.

Fig.\,\ref{Fig1}(b) shows a typical surface obtained for 
CaK(Fe$_{0.95}$Ni$_{0.05}$)$_4$As$_4$ which resembles surfaces
obtained
 in pure CaKFe$_4$As$_4$\cite{PhysRevB.97.134501}. We identify
atomically flat areas over a scanning window several $\mu$m in size, separated
by atomic size trenches (black lines in Fig.\,\ref{Fig1}(b)).  Fig.\,\ref{Fig1}(c) displays the
tunneling conductance $G=dI/dV$. The superconducting gap manifests as the usual, strong reduction of the tunneling conductance for
voltages of order of a few mV, which disappears at about $T_c$. The zero bias density of states is finite and the coherence peaks are
strongly smeared. Under magnetic fields we observed a disordered hexagonal vortex lattice (lower left inset of Fig.\,\ref{Fig1}(c) and Ref.\cite{SupplementMat}). To estimate the superconducting gap at zero field in
CaK(Fe$_{0.95}$Ni$_{0.05}$)$_4$As$_4$, we construct a superconducting density
of states $N(E)$ allowing for a large distribution of values of the superconducting gap (details in Ref.\,\cite{SupplementMat}) which gives the tunneling conductance as solid lines in the main panel of Fig.\,\ref{Fig1}(c). The lower right inset of Fig.\,\ref{Fig1}(c) shows the energy at
which the derivative of $N(E)$, $\frac{dN}{dE}$, has a maximum. We obtain $1.5$ meV at low temperatures, which is very similar to the gap value estimated from
$T_c\approx10$ K, $\Delta\approx 1.2$ meV. This value decreases with temperature as shown in the lower right panel of Fig.\,\ref{Fig1}(c). These results are completely different from those observed in pure CaKFe$_4$As$_4$, where a two-gap structure with a few states at the Fermi level is found\cite{PhysRevB.97.134501,PhysRevB.95.100502}.

\begin{figure*}
      \includegraphics[width=\textwidth]{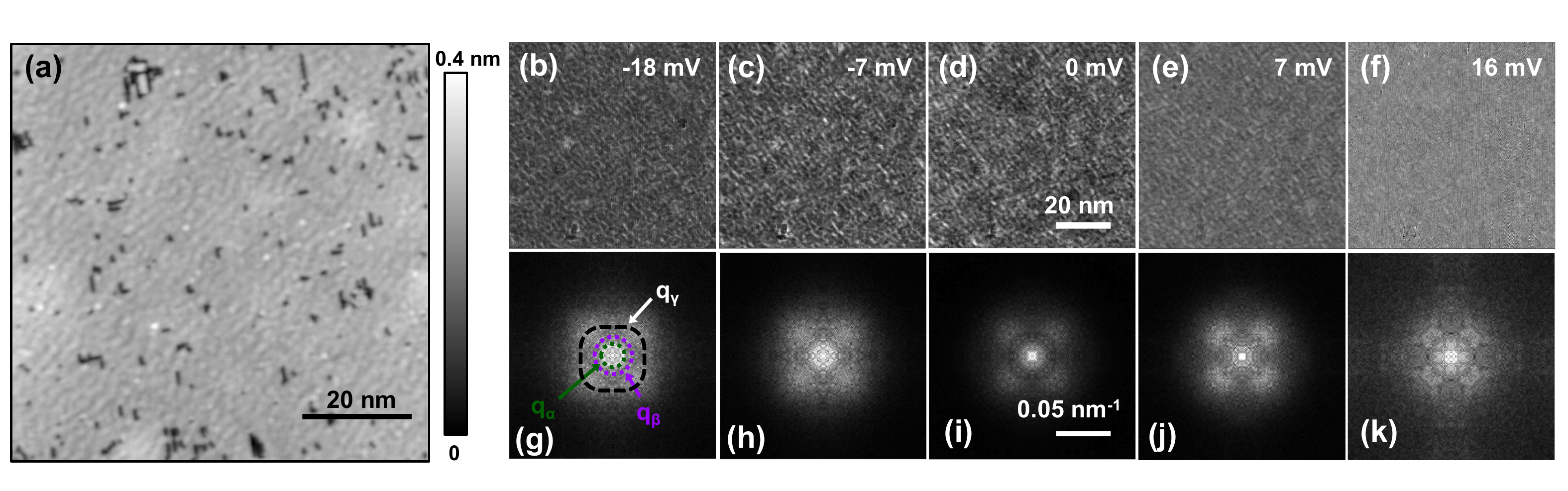}
	\caption{(a) Topography of the area where we have made the quasiparticle interference experiment shown in (b-k). The color scale bar is given in the bottom right and the gray scale by the bar at the right. The image has been taken at a bias voltage of 30 mV and a current of 1 nA at zero magnetic field and at 0.3 K. (b-f) Tunneling conductance as a function of the position for a few representative bias voltages (given in each panel). The lateral scale bar is provided in (d). (g-k) Fourier transform, symmetrized taking into account the in-plane square lattice, of (b-f) shown in the first Brillouin zone. In (g) we mark the outer main scattering vector (black dashed circle) as well as the two inner scattering vectors (purple and green dashed circles). The lateral scale bar is given in (i) and grey scale goes from low (black) to large (white) scattering intensity.}
	\label{Fig_QPI}
	\end{figure*}

When zooming into a small region we observe strong electronic
scattering due to defects. The field of view shown in the topographic constant current image of Fig.\,\ref{Fig_QPI}(a)) is atomically flat. There are atomic size defects (black spots) and there is a wavy background. We can then build tunneling conductance maps $G(V,x,y)$ at each point $(x,y)$ of the field of view. A representative example is shown in Figs.\,\ref{Fig_QPI}(b-f) for a few bias voltages $V$. $G(V)$ is quite homogeneous and does not change much close to atomic size defects but we can identify clearly a wavy background whose wavelength changes with V.
The contribution of scattered electrons to the
$G(V)$ is proportional to the densities of states of initial and final states, i.e. the
joint density of states, and the scattering wavelength is equal to the difference
$q$ between initial and final scattering wavevectors\cite{Hoffman_2011,PhysRevB.97.134501}. The Fourier transform of the
tunneling conductance images is shown in
Fig.\,\ref{Fig_QPI}(g-k). We identify three main scattering vectors,
$q_{\alpha}$, $q_\beta$ and $q_\gamma$.  The largest scattering vector, $q_\gamma$, is slightly anisotropic, being larger along the $\overline{\Gamma}$-$\overline{X}$
direction than along  $\overline{\Gamma}$-$\overline{M}$ (notation of Brillouin zone directions follows the one proposed for pnictide superconductors in Ref.\,\cite{doi:10.1002/andp.201000149}). The Fourier amplitude at the three
scattering vectors decreases close to the Fermi level due to the opening of the
superconducting gap (Figs.\,\ref{Fig_QPI}(i)). 

\begin{figure}[h]
      \includegraphics[width=0.45\textwidth]{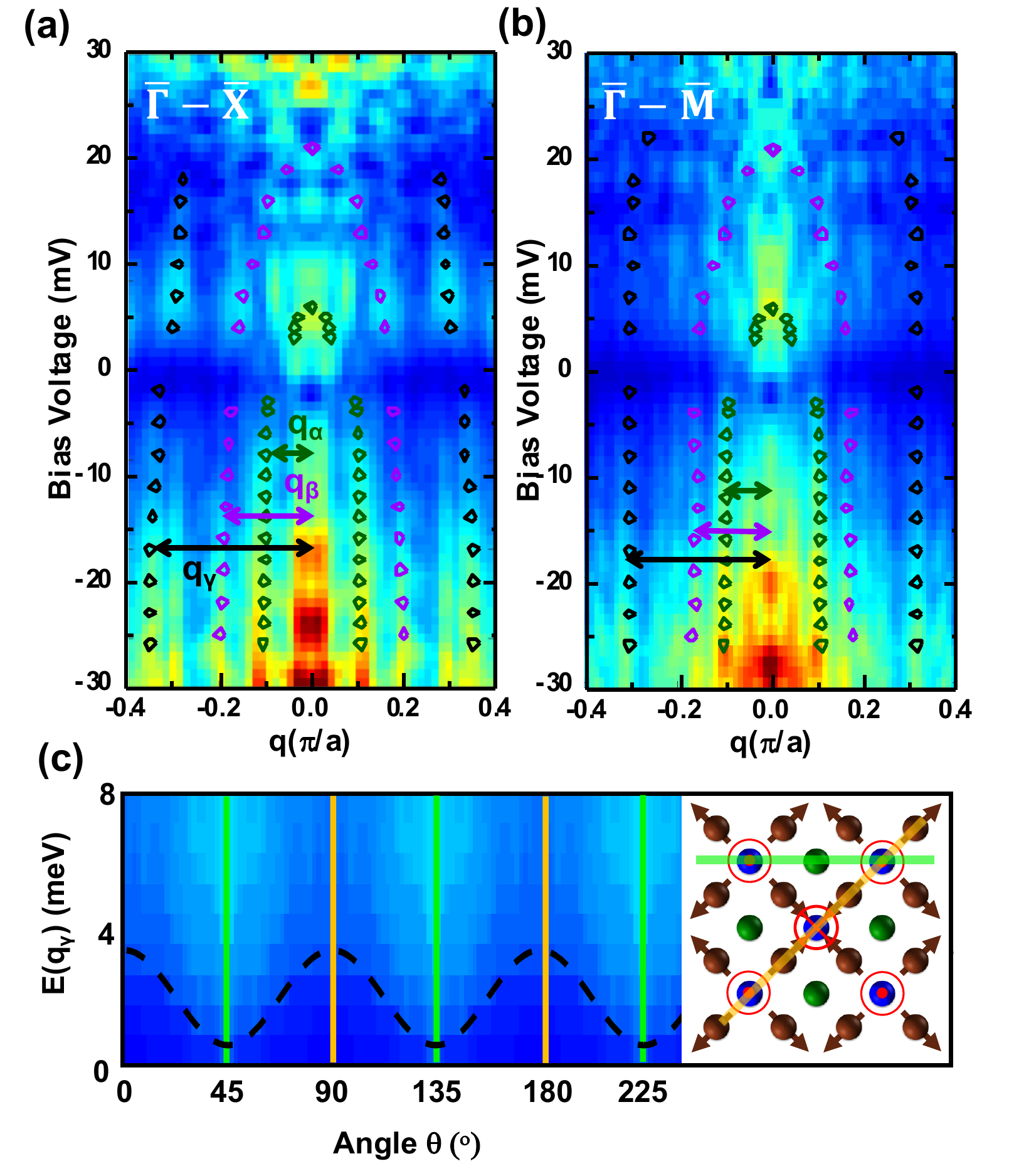}
	\caption{(a) Scattering intensity for the two main
	symmetry directions, $\overline{\Gamma}$-$\overline{X}$ (left panel) and $\overline{\Gamma}$-$\overline{M}$ (right panel). Open circles mark the evolution of
	the scattering vectors with bias voltage. Scattering vectors
	$q_{\alpha}$, $q_{\beta}$ and $q_{\gamma}$ are shown in black, violet and green. Color scale goes from low (blue) to high (red) scattering intensity.
	(b) Scattering intensity as a
	function of the angle with respect to the in-plane a axis for energies
	close to the Fermi level in $q_\gamma$. Color scale goes from blue (low intensity) to cyan (high intensity). The vertical light green and orange lines highlight the $\overline{\Gamma}$-$\overline{M}$ and $\overline{\Gamma}$-$\overline{X}$ directions respectively. The black dashed curve is a guide to the eye. The  right inset shows a schematic representation of the lattice, with Fe 	atoms in brown (and their spins represented by arrows), As1 in blue and As2 in green and the main symmetry directions as light green and orange lines.}
	\label{Fig_QPI2}
	\end{figure}

When plotting the bias voltage dependence of the scattering
pattern along the high symmetry directions $\overline{\Gamma}$-$\overline{X}$ and $\overline{\Gamma}$-$\overline{M}$
(Figs.\,\ref{Fig_QPI2}(a) and (b) respectively), we observe that all scattering vectors  $q$
decrease in size when increasing
the bias voltage above the Fermi level. The qualitative behavior is very similar for both high symmetry directions, although the values of $q_{\gamma}$ are slightly larger for $\overline{\Gamma}$-$\overline{X}$ than for $\overline{\Gamma}$-$\overline{M}$. The reduction
of the intensity inside the superconducting gap is band dependent. The
superconducting gap is most clearly observed for the largest scattering vector $q_\gamma$. 

In Fig.\,\ref{Fig_QPI2}(c) we show the scattering
intensity around zero bias and at $q_\gamma$ as a function of the angle, with $\theta=0^{\circ}$ for $\overline{\Gamma}$-$\overline{X}$ and $\theta=45^{\circ}$ for $\overline{\Gamma}$-$\overline{M}$. We find a 
 fourfold modulation of the superconducting density of states
  which is not present in the stoichiometric
 compound and follows the symmetry of the SVC\cite{PhysRevB.97.134501,PhysRevB.95.100502}. 
 The superconducting gap is larger along the direction where the
 hyperfine field on the As1 sites cancels ($\overline{\Gamma}$-$\overline{X}$, orange lines in Fig.\,\ref{Fig_QPI2}(c)), whereas it is smaller when the hyperfine field remains finite ($\overline{\Gamma}$-$\overline{M}$, green lines in Fig.\,\ref{Fig_QPI2}(c)), suggesting a competing scenario
 between superconductivity and magnetism. We will analyze this
 observation further below.

\begin{figure}
      \includegraphics[width=0.45\textwidth]{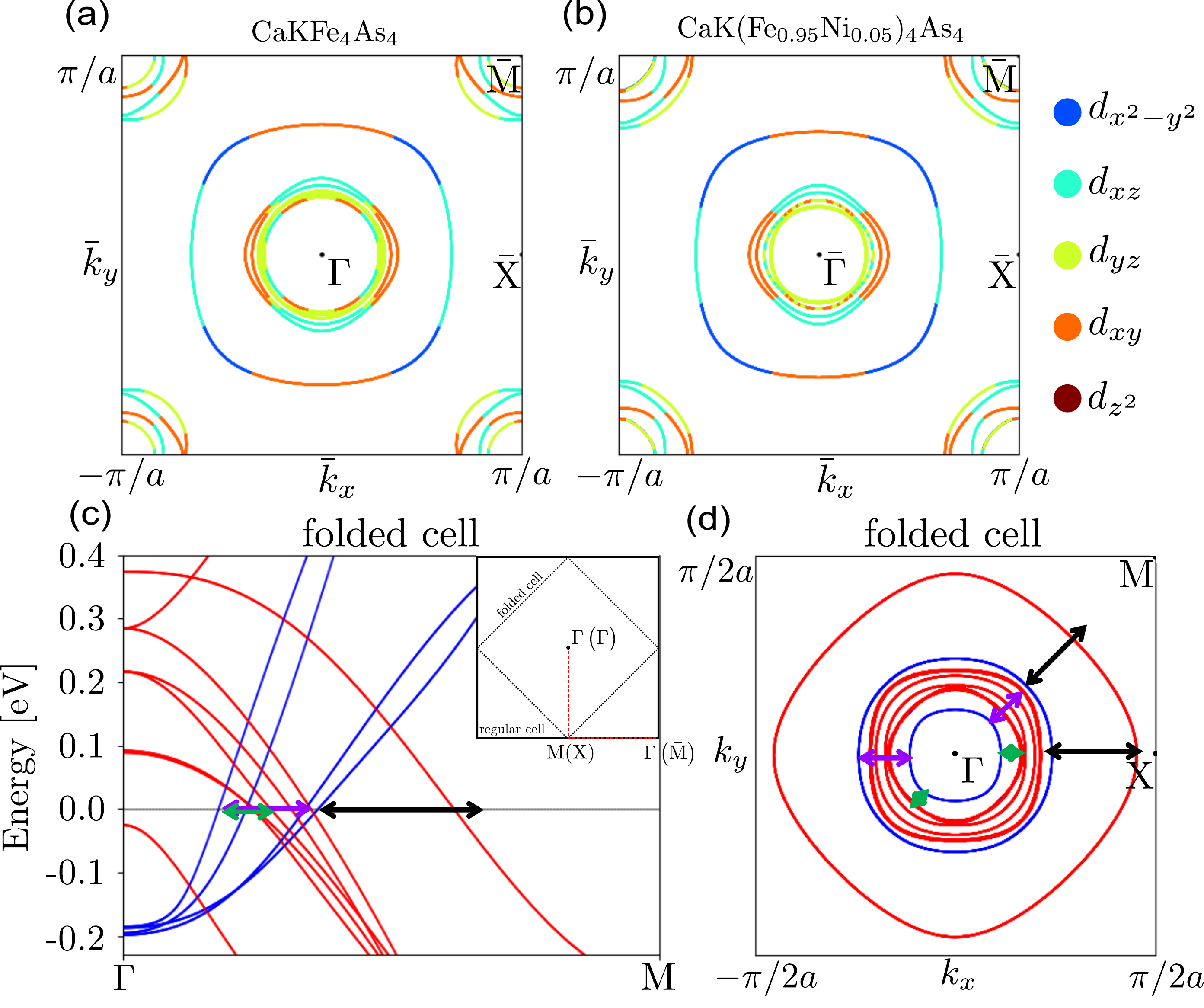}
	\caption{Fermi surface of (a) pure CaKFe$_4$As$_4$ 
	and (b) CaK(Fe$_{0.95}$Ni$_{0.05}$)$_4$As$_4$ 
	in the paramagnetic phase, obtained as described in the text.
	(c,d) Band structure and  Fermi surface of CaK(Fe$_{0.95}$Ni$_{0.05}$)$_4$As$_4$ in the folded AFM Brillouin zone. Folded bands are
	shown in blue.
	Black, violet and green arrows are the main scattering vectors shown in Fig.\ref{Fig_QPI2}. Convention for the names of directions of folded and unfolded Brillouin zones (inset of (c)) follows Ref.\,\cite{doi:10.1002/andp.201000149}.}
	\label{Fig_band structure}
	\end{figure}

In what follows we investigate the origin of the
three scattering vectors identified in Fig.\,\ref{Fig_QPI2}(a,b).
We have calculated the electronic 
structure of CaKFe$_4$As$_4$  and 
CaK(Fe$_{0.95}$Ni$_{0.05}$)$_4$As$_4$ in the tetragonal
paramagnetic phase  within density functional theory
as described in Ref.\,\cite{SupplementMat}. The effect of Ni doping has been taken into
account with the Virtual Crystal Approximation (VCA). In 
Fig.\,\ref{Fig_band structure}(a) and (b) we show the respective
Fermi surfaces. As expected, upon Ni doping
 the inner hole
pockets slightly shrink
in CaK(Fe$_{0.95}$Ni$_{0.05}$)$_4$As$_4$ (Fig.\,\ref{Fig_band
structure}(b)) as compared to the pure compound
(Fig.\,\ref{Fig_band structure}(a)) with  the overall structure of the
Fermi surface remaining similar. 
Our measurements (e.g. Fig.\,\ref{Fig_QPI2}(a,b))
show, however, that the scattering pattern is very
different. In
CaKFe$_4$As$_4$~\cite{PhysRevB.97.134501} the scattering pattern consists of a single scattering vector, associated to interband
scattering between two hole bands centered at the Brillouin zone that increases
strongly in size when increasing the bias voltage. In
CaK(Fe$_{0.95}$Ni$_{0.05}$)$_4$As$_4$ there are three vectors whose size
decreases much less drastically above the Fermi level.
The SVC magnetic order invokes a folding of the band structure
along the AFM wavevector due to the doubling of the unit cell 
(see inset in Fig. \,\ref{Fig_band structure}(c)). We assume that folding is the main consequence of the SVC in the bandstructure. The
 folded electron bands are shown in Fig. \,\ref{Fig_band structure}(c) and the
 Fermi surface in Fig. \,\ref{Fig_band structure}(d)).  
The  bands at the edges of the unfolded Brillouin zone are now
centered around $\overline{\Gamma}$, providing a clearly defined set of bands
coexisting in the same Brillouin zone region as the hole pockets
centered at $\overline{\Gamma}$.
In the calculated bands we identify three scattering vectors between hole and
electron bands whose size corresponds to the observed $q_{\alpha}$, $q_\beta$
and $q_\gamma$ vectors (arrows in Figs.\,\ref{Fig_band structure}(c,d)). Their
value decreases with increasing bias voltage as is also
found  experimentally
Fig.\,\ref{Fig_QPI2}(a,b). Thus, the reconstructed Fermi surface provides an
accurate description of the band structure of
CaK(Fe$_{0.95}$Ni$_{0.05}$)$_4$As$_4$.

We discuss now the observed 
fourfold modulation of the superconducting gap in
CaK(Fe$_{0.95}$Ni$_{0.05}$)$_4$As$_4$ (Fig.\,\ref{Fig_QPI2}(c)) which is not
present in the stoichiometric
compound\cite{PhysRevB.97.134501,PhysRevB.95.100502}.
NMR experiments, M\"{o}ssbauer spectroscopy and muon spin rotation/relaxation
studies have shown evidence for the coexistence between superconductivity and
the magnetic order in
CaK(Fe$_{0.95}$Ni$_{0.05}$)$_4$As$_4$\cite{Ding17,bud2018coexistence,khasanov2020magnetism}.
It has been also reported that, similar to what is found in 122 compounds
Ba$_{1-x}$M$_{x}$Fe$_2$As$_2$ with M = Co, Ni and
Rh\cite{fernandes2010unconventional,kreyssig2010suppression,luo2012coexistence}
and Ba(Fe$_{1-x}$K$_x$)$_2$As$_2$\cite{munevar2013superconductivity}, the
ordered magnetic moment is gradually suppressed when entering in the
superconducting phase, suggesting that superconductivity and magnetism are
competing for the same electrons in the iron-based
superconductors\cite{kreyssig2018antiferromagnetic,bud2018coexistence,khasanov2020magnetism,bud2018coexistence,machida1981spin}.
Our results show that this competition is also associated with the development of a strongly anisotropic superconducting gap.

The SVC phase is the only magnetic phase of pnictide superconductors where glide symmetry is broken within the unit cell. The relation between glide symmetry and superconductivity is not direct, because the coherence length is larger than the unit cell size. However, it may lead to a spin-current density wave, or $d$-density wave with increasing temperature or disorder\cite{PhysRevB.93.014511}. The chiral properties of a spin-current density wave are connected to a pattern of currents inside the unit cell. The $d$-density wave has been suggested to be related to situations with hidden order parameters, such as the low temperature ordered phase of URu$_2$Si$_2$ or the pseudogap in the cuprates\cite{PhysRevB.66.224505,PhysRevB.39.2940,PhysRevB.63.094503}. By contrast to usual magnetic fluctuations, which peak at horizontal directions on the Brillouin zone and favor repulsive interactions, fluctuations related to SVC peak at the corners of the Brillouin zone are attractive\cite{PhysRevB.93.014511}. Thus, there are in principle no expected modifications of $s\pm$ pairing in the SVC. However, the absence of local inversion symmetry, with the associated potential spin current $d$-density wave plaquette pattern is likely to have a strong influence on the superconducting properties and might produce the fourfold nodeless anisotropic gap observed here\cite{PhysRevB.66.224505,PhysRevB.39.2940,PhysRevB.63.094503,PhysRevB.93.014511}. London penetration depth measurements in pure and electron irradiated crystals of CaK(Fe$_{0.95}$Ni$_{0.05}$)$_4$As$_4$ have suggested the presence of an anisotropic superconducting gap with $s\pm$ symmetry which is more sensitive to disorder than in the stoichiometric compound \cite{teknowijoyo2018robust}. This can also explain the observation of a finite density of states at zero bias in CaK(Fe$_{0.95}$Ni$_{0.05}$)$_4$As$_4$. 

In conclusion, we have measured the spatial dependence of the tunneling conductance
in the SVC state of CaK(Fe$_{0.95}$Ni$_{0.05}$)$_4$As$_4$ and report direct
evidence for a strong mutual influence between superconductivity and SVC
antiferromagnetic order. Quasiparticle interference measurements 
supported by band structure calculations demonstrate a Fermi surface reconstruction and anisotropic pairing through an
in-plane fourfold modulation of the superconducting gap. The comparison to
CaKFe$_4$As$_4$, where there is no antiferromagnetic order and the
superconducting gap shows no in-plane anisotropy, strongly suggests that the SVC
antiferromagnetic state is responsible for the anisotropic pairing in
CaK(Fe$_{0.95}$Ni$_{0.05}$)$_4$As$_4$.

\section{Acknowledgments}
This work was supported by the Spanish Research State Agency (FIS2017-84330-R,
RYC-2014-15093, CEX2018-000805-M), by the European Research Council PNICTEYES
grant agreement 679080 and by EU program Cost CA16218 (Nanocohybri), by the
Comunidad de Madrid through program NANOMAGCOST-CM (Program No.
S2018/NMT-4321)
and by the Deutsche Forschungsgemeinschaft (DFG, German Research Foundation)
through TRR 288 - 422213477 (project A05). The research was supported by the U.S. Department of Energy (DOE), Office of Basic Energy Sciences, Division of Materials Sciences and Engineering. Ames Laboratory is operated for the U.S. DOE by the Iowa State University under Contract No. DE-AC02-07CH11358. WRM was supported by the Gordon and Betty Moore Foundation’s EPiQS Initiative through Grant GBMF4411. 
We acknowledge SEGAINVEX at UAM for design and
construction of cryogenic equipment and the computational resources of the computer center of the Goethe University Frankfurt. We also thank R. \'Alvarez Montoya, S. Delgado and J.M. Castilla for technical support. 

%

\end{document}